\documentclass[12pt]{article}
\usepackage{cite}
\usepackage{a4}
\usepackage{ulem}
\usepackage{color}
\usepackage{epsfig}
\usepackage{graphicx}
\usepackage{subfigure}
\usepackage{ulem}
\usepackage{amsfonts}
\usepackage{epsfig,bm}
\usepackage{graphicx}
\usepackage{comment}
\newcommand{\be}{\begin{equation}}
\newcommand{\ee}{\end{equation}}
\newcommand{\ba}{\begin{eqnarray}}
\newcommand{\ea}{\end{eqnarray}}
\newcommand{\nn}{\nonumber\\}
\begin{document}
\begin{titlepage}
\title{State-space Geometry, Statistical Fluctuations \\ and \\ Black Holes in
String Theory}
\author{}
\date{Stefano Bellucci$^{a}$ \thanks{\noindent bellucci@lnf.infn.it} \
and\ Bhupendra Nath Tiwari$^{a}$ \thanks{\noindent bntiwari.iitk@gmail.com}\\
\vspace{0.5cm}
$^{a}$INFN-Laboratori Nazionali di Frascati\\
Via E. Fermi 40, 00044 Frascati, Italy.}
\maketitle
\abstract{We study the state-space geometry of various extremal
and nonextremal black holes in string theory. From the notion of
the intrinsic geometry, we offer a state-space perspective to the
black hole vacuum fluctuations. For a given black hole entropy,
we explicate the intrinsic geometric meaning of the statistical
fluctuations, local and global stability conditions and long range
statistical correlations. We provide a set of physical motivations
pertaining to the extremal and nonextremal black holes,
\textit{viz.}, the meaning of the chemical geometry and physics of
correlation. We illustrate the state-space configurations for
general charge extremal black holes. In sequel, we extend our
analysis for various possible charge and anticharge nonextremal
black holes. From the perspective of statistical fluctuation
theory, we offer general remarks, future directions and open
issues towards the intrinsic geometric understanding of the vacuum
fluctuations and black holes in string theory.}

\vspace{0.50cm}

\end{titlepage}
\section{Introduction}
In this paper, we study statistical properties of the charged
anticharged black hole configurations in string theory.
Specifically, we illustrate that the components of the vacuum
fluctuations define a set of local pair correlations against the
parameters, \textit{e.g.}, charges, anticharges, mass and angular momenta.
Our consideration follows from the notion of the thermodynamic
geometry, mainly introduced by Weinhold \cite{wein1,wein2} and
Ruppeiner \cite{rup1,rup11, rup12, rup2,rup21,rup22,
RuppeinerPRD78}. Importantly, this framework provides a simple
platform to geometrically understand the statistical nature of
local pair correlations and underlying structures pertaining to
the vacuum phase transitions. In diverse contexts, the state-space
geometric perspective offers an understanding of the phase structures
of mixtures of gases, black hole configurations
\cite{bnt,bnt1,bnt2,bnt3,bnt5,bnt6,bnt7,
bnt8,bnt9,bnt10,aman,aman1,aman2,aman2a,aman3,aman4,aman5}, generalized
uncertainty principle \cite{bntoo}, strong interactions, e.g., hot
QCD \cite{bntsbvc}, quarkonium configurations \cite{bullquark},
and some other systems, as well.

The main purpose of the present article is to consider the
state-space properties of various possible extremal and
nonextremal black holes in string theory, in general. String
theory \cite{Witten}, as the most promising framework to
understand all possible fundamental interactions, celebrates the
physics of black holes, in both the zero and the nonzero
temperature domains. Our consideration hereby plays a crucial role
in understanding the possible phases and stability of the string
theory vacua. A further motivation follows from the consideration
of the string theory black holes. Namely, $\mathcal N=2$
supergravity arises as a low energy limit of the Type II string theory
solution, admitting extremal black holes with the zero Hawking
temperature and a nonzero macroscopic attractor entropy.

A priori, the entropy depends on a large number of scalar moduli
arising from the compactification of the 10 dimension theory down
to the 4 dimensional physical spacetime. This involves a 6
dimensional compactifying manifold. Interesting string theory
compactifications involve $T^6$, $K_3 \times T^2$ and Calabi-Yau
manifolds. The macroscopic entropy exhibits a fixed point behavior
under the radial flow of the scalar fields. In such cases, the
near horizon geometry of an extremal black hole turns out to be an
$AdS_2 \times S^2 $ manifold which describes the Bertotti-Robinson
vacuum associated with the black hole. The area of the black hole
horizon $A$ and thus the macroscopic entropy \cite{9508072v3,
9602111v3, new1, new2,bfm1,bfm2,bfgm1,bfmy,0702019v1, bskm,
08051310,bfgm2} is given as $S_{macro}= \pi \vert Z_{\infty}
\vert^2$. This is known as the Ferrara-Kallosh-Strominger
attractor mechanism, which as the macroscopic consideration,
requires a validity from the microscopic or statistical basis of
the entropy. In this concerns, there have been various
investigations on the physics of black holes, e.g., horizon
properties \cite{0505122v2,0411255}, counting of black hole
microstates \cite{0409148,0507014v1,0502157v4}, spectrum of
half-BPS states in $\mathcal{N} = 4$ supersymmetric string theory
\cite{0504005} and fractionation of branes \cite{0611330}. From
the perspective of the fluctuation theory, our analysis is
intended to provide the nature of the statistical structures of
the extremal and nonextremal black hole configurations. The
attractor configurations exist for the extremal black holes, in
general. However, the corresponding nonextremal configurations
exist in the throat approximation. In this direction, it is worth
mentioning that there exists an extension of Sen entropy function
formalism for $ D_1D_5 $ and $ D_2D_6NS_5 $ non-extremal
configurations \cite{GarousiGhodsiCai,
GarousiGhodsiCai1,GarousiGhodsiCai2}. In the throat approximation,
these solutions respectively correspond to Schwarzschild black
holes in $ AdS_3 \times S^3 \times T^4 $ and $ AdS_3 \times S^2
\times S^1 \times T^4 $. In relation with the intrinsic
state-space geometry, we shall explore the statistical
understanding of the attractor mechanism and the moduli space
geometry, and explain the vacuum fluctuations of the black brane
configurations.

In this paper, we consider the state-space geometry of the
spherical horizon topology black holes in four spacetime
dimensions. These configurations carry a set of electric magnetic
charges $(q_i, p_i)$. Due to the consideration of Strominger and
Vafa \cite{9601029v2}, these charges are associated to an ensemble
of weakly interacting D-branes. Following the Refs.
\cite{9512078v1,9601029v2,MSW,9602043v2,
9603109v1,9603195v1,9603061v2}, it turns out that the charges
$(q_i, p_i)$ are proportional to the number of electric and
magnetic branes, which constitute the underlying ensemble of the
chosen black hole. In the large charge limit, \textit{viz.}, when
the number of such branes becomes large, we have treated the
logarithm of the degeneracy of states of the statistical
configuration as the Bekenstein-Hawking entropy of the associated
string theory black holes. For the extremal black holes, the
entropy is described in terms of the number of the constituent
D-branes. For example, the two charge extremal configurations can
be examined in terms of the winding modes and the momentum modes
of an excited string carrying $ n_1 $ winding modes and $ n_p $
momentum modes. Correspondingly, the state-space geometry of the
non-extremal black holes are described by adding energy to the
extremal D-branes configurations. This renders as the contribution
of the clockwise and anticlockwise momenta in the Kaluza-Klein
scenarios and that of the antibrane charges in general to the
black hole entropy.

From the perspective of black hole thermodynamics, we describe the
structure of the state-space geometry of four dimensional extremal
and nonextremal black holes in a given duality frame. Thus, when
we take arbitrary variations over the charges $(q_i,p_i)$ on the
electric and magnetic branes, the underlying statistical
fluctuations are described by only the numbers of the constituent
electric and magnetic branes. From the perspective of the
intrinsic state-space geometry, if one pretends that the notion of
statistical fluctuations applies to intermediate regimes of the
moduli space, then the attractor horizon configurations require an
embedding to the higher dimensional intrinsic Riemanian manifold.
Physically, such a higher dimensional manifold can be viewed as a
possible blow up of the attractor fixed point phase-space to a
non-trivial moduli space. From the perspective of thermodynamic
Ruppenier geometry, we have offered future directions and open
issues in the conclusion. We leave the explicit consideration of
these matters open for further research.

In section 2, we define the general notion of vacuum fluctuations.
This offers the physical meaning of the state-space geometry. In
section 3, we provide a brief review of statistical fluctuations.
In particular, for a given black hole entropy, we firstly
explicate the statistical meaning of state-space surface, and then
offer the general meaning of the local and global stability
conditions and long range statistical correlations. In section 4,
we provide a set of physical motivations pertaining to the
extremal and nonextremal black holes, the meaning of Wienhold
chemical geometry and the physics of correlation. In section 5, we
consider state-space configurations pertaining to the extremal
black holes and explicate our analysis for the two and three
charge configurations. In section 6, we extend the above analysis
for the four, six and eight charge anticharge nonextremal black
holes. Finally, section 7 provides general remarks, conclusion 
and outlook, and future directions and open issues
towards the application of string theory.
\section{Definition of State-space Geometry}
Considering the fact that the black hole configurations in string
theory introduce the notion of vacuum, it turns out for any
thermodynamic system, that there exist equilibrium thermodynamic
states given by the maxima of the entropy. These states may be
represented by points on the state-space. Along with the laws of
the equilibrium thermodynamics, the theory of fluctuations leads
to the intrinsic Riemannian geometric structure on the space of
equilibrium states \cite{rup22,RuppeinerPRD78}. The invariant
distance between two arbitrary equilibrium states is inversely
proportional to the fluctuations connecting the two states. In
particular, a less probable fluctuation means that the states are
far apart. For a given set of states $\{X_i\}$, the state-space
metric tensor is defined by
\begin{eqnarray} \label{metricdef}
g_{ij}(X)&=& -\partial_i \partial_j S( X_1, X_2,\dots, X_n)
\end{eqnarray}
A physical motivation of Eq.(\ref{metricdef}) can be given as
follows. Up to the second order approximation, the Taylor
expansion of the entropy $S( X_1, X_2,\dots, X_n)$ yields
\begin{eqnarray}
S- S_0 = -\frac{1}{2} \sum_{i=1}^n  g_{ij}\Delta X^i \Delta X^j,
\end{eqnarray}
where
\begin{eqnarray}
g_{ij}:= -\frac{\partial^2 S( X_1, X_2,\dots, X_n) }{\partial X^i
\partial X^j}=  g_{ji}
\end{eqnarray}
is called the state-space metric tensor. In the present
investigation, we consider the state-space variables $\{ X_1,
X_2,\dots, X_n \}$ as the parameters of the ensemble of the
microstates of the underlying microscopic configuration (e.g.
conformal field theory \cite{maldacena}, black hole conformal
field theory, \cite{0412322}, hidden conformal field theory
\cite{hidden1,hidden2}, etc.), which defines the corresponding
macroscopic thermodynamic configuration. Physically, the
state-space geometry can be understood as the intrinsic Riemannian
geometry involving the parameters of the underlying microscopic
statistical theory. In practice, we shall consider the variables
$\{ X_1, X_2,\dots, X_n \}$ as the parameters, \textit{viz.},
charges, anticharges and others if any, of the corresponding low
energy limit of the string theory, e.g., $\mathcal{N} = 2$
supergravity. In the limit, when all the variables, \textit{viz.},
$\{ X_1, X_2,\dots, X_n \}$ are thermodynamic, the state-space
metric tensor Eq.(\ref{metricdef}) reduces to the corresponding
Ruppenier metric tensor. In the discrete limit, the relative
co-ordinates $\Delta X^i$ are defined as $\Delta X^i:= X^i-
X^i_0$, for given $\{X^i_0\} \in M_n$. In the Gaussian
approximation, the probability distribution has the following form
\begin{eqnarray}
P( X_1, X_2,\dots, X_n)= A\ \exp(-\frac{1}{2} g_{ij}\Delta X^i
\Delta X^j)
\end{eqnarray}
With the normalization
\begin{eqnarray}
\int \prod_i dX_i  P( X_1, X_2,\dots, X_n)= 1,
\end{eqnarray}
we have the following probability distribution
\begin{eqnarray}
P( X_1, X_2,\dots, X_n)= \frac{\sqrt{g(X)}}{(2 \pi)^{n/2}}
exp(-\frac{1}{2} g_{ij} dX^i \otimes dX^j),
\end{eqnarray}
where $g_{ij}$ now, in a strict mathematical sense, is properly
defined as the inner product $g(\frac{\partial}{\partial X^i},
\frac{\partial}{\partial X^j})$ on the corresponding tangent space
$T(M_n) \times T(M_n)$. In this connotation, the determinant of
the state-space metric tensor
\begin{eqnarray}
g(X):= \Vert g_{ij} \Vert
\end{eqnarray}
can be understood as the determinant of the corresponding matrix
$[g_{ij}]_{n \times n}$. For a given state-space manifold $(M_n,
g)$, we shall think of $\{dX^i \}$ as the basis of the cotangent
space $T^{\star}(M_n)$. In the subsequent analysis, by taking an
account of the fact that the physical vacuum is neutral, we shall
choose $X^i_0=0$.
\section{Statistical Fluctuations}
\subsection{Black Hole Entropy}
As a first exercise, we have illustrated thermodynamic state-space
geometry for the two charge extremal black holes with electric
charge $q$ and magnetic charge $p$. The next step has thence been
to examine the thermodynamic geometry at an attractor fixed
point(s) for the extremal black holes as the maxima of their
macroscopic entropy $S(q,p)$. Later on, the state-space geometry
of nonextremal counterparts has as well been analyzed. In this
investigation, we demonstrate that the state-space correlations of
nonextremal black holes modulate relatively more swiftly to an
equilibrium statistical basis than those of the corresponding
extremal solutions.
\subsection{State-space Surface}
The Ruppenier metric on the state-space $(M_2,g)$ of two charge
black hole is defined by
\begin{eqnarray}
g_{qq}=- \frac{\partial^2 S(q,p)}{\partial q^2}, \ \ g_{qp}=-
\frac{\partial^2 S(q,p)}{{\partial q}{\partial p}} , \ \ g_{pp}=-
\frac{\partial^2 S(q,p)}{\partial p^2}
\end{eqnarray}
Subsequently, the components of the state-space metric tensor are
associated to the respective statistical pair correlation
functions. It is worth mentioning that the co-ordinates on the
state-space manifold are the parameters of the microscopic
boundary conformal field theory which is dual the black hole
space-time solution. This is because the underlying state-space
metric tensor comprises of the Gaussian fluctuations of the
entropy which is the function of the number of the branes and
antibranes. For the chosen black hole configuration, the local
stability of the underlying statistical system requires both
principle minors to be positive. In this se-up, the diagonal
components of the state-space metric tensor, viz., $\{ g_{x_ix_i}
\mid x_i= (n,m)\}$ signify the heat capacities of the system. This
requires that the diagonal components of the state-space metric
tensor
\begin{eqnarray}
g_{x_ix_i} &>& 0, \ i= \ n, m
\end{eqnarray}
be positive definite. In this investigation, we discuss the
significance of the above observation for the eight parameter
non-extremal black brane configurations in string theory. From the
notion of the relative scaling property, we shall demonstrate the nature
of the brane-brane pair correlations. Namely, from the perspective of the
intrinsic Riemannian geometry, the stability properties of the
eight parameter black branes are examined from the positivity of
the principle minors of the space-state metric tensor. For the
Gaussian fluctuations of the two charge equilibrium statistical
configurations, the existence of a positive definite volume form
on the state-space manifold $(M_2(R),g)$ imposes such a global
stability condition. In particular, the above configuration leads
to a stable statistical basis, if the determinant of the
state-space metric tensor
\begin{eqnarray}
\Vert g \Vert &= &S_{nn}S_{mm}- S_{nm}^2
\end{eqnarray}
remains positive. Indeed, for the two charge black brane configurations,
the geometric quantities corresponding to the underlying
state-space manifold elucidates typical features of the Gaussian
fluctuations about an ensemble of equilibrium brane microstates.
In this case, we see that the Christoffel connections on the 
$(M_2,g)$ are defined by
\begin{eqnarray}
\Gamma_{ijk} = g_{ij,k} + g_{ik,j}- g_{jk,i}
\end{eqnarray}
The only nonzero Riemann curvature tensor is
\begin{eqnarray}
R_{qpqp}= \frac{N}{D},
\end{eqnarray}
where
\begin{eqnarray}
N&:=& S_{pp}S_{qqq}S_{qpp} + S_{qp}S_{qqp}S_{qpp} \nonumber \\ &&+
S_{qq}S_{qqp}S_{ppp} -S_{qp}S_{qqq}S_{ppp} \nonumber \\ &&-
S_{qq}S_{qpp}^2- S_{pp}S_{qqp}^2
\end{eqnarray}
and
\begin{eqnarray}
D:=  (S_{qq}S_{pp}- S_{qp}^2)^{2}
\end{eqnarray}
The scalar curvature and the corresponding $R_{ijkl}$ of an
arbitrary two dimensional intrinsic state-space manifold
$(M_2(R),g)$ may be given as
\begin{eqnarray}
R(q,p)=\frac{2}{\Vert g \Vert}R_{qpqp}(q,p)
\end{eqnarray}
\subsection{Stability Conditions}
For a given set of variables $\lbrace X^1, X^2, \ldots, X^n
\rbrace$, the local stability of the underlying state-space
configuration demands the positivity of the heat capacities
\begin{eqnarray}
\lbrace g_{ii}(X^i) > 0; \ \forall i= 1, 2, \ldots, n \rbrace
\end{eqnarray}
Physically, the principle components of the state-space metric
tensor $\lbrace g_{ii}(X^i) \ \vert \ i= 1, 2, \ldots, n \rbrace$
signify a set of definite heat capacities (or the related
compressibilities), whose positivity apprises that the black hole
solution comply an underlying, locally in equilibrium, statistical
configuration. Notice further that the positivity of principle
components is not sufficient to insure the global stability of the
chosen configuration and thus one may only achieve a locally in
equilibrium configuration. In fact, the global stability condition
constraint over the allowed domain of the parameters of black hole
configurations requires that all the principle components and all
the principle minors of the metric tensor must be strictly
positive definite \cite{rup2}. The above stability conditions
require that the following set of equations must be simultaneously
satisfied
\begin{eqnarray}
p_0&:=& 1,  \nonumber \\
p_1&:=& g_{11} > 0, \nonumber \\
p_2&:=&  \left \vert\begin{array}{rr}
    g_{11} & g_{12}  \\
     g_{12} & g_{22}  \\
\end{array} \right \vert > 0, \nonumber \\
p_3&:=&  \left \vert\begin{array}{rrr}
    g_{11} & g_{12} & g_{13} \\
     g_{12} & g_{22} & g_{23} \\
     g_{13} & g_{23} & g_{33} \\
\end{array} \right \vert > 0, \nonumber \\
\vdots \nonumber \\
p_n&:=& \Vert g \Vert > 0
\end{eqnarray}
\subsection{Long Range Correlations}
The thermodynamic scalar curvature of the state-space manifold is
proportional to the correlation volume. Physically, the scalar
curvature signifies the interaction(s) of the underlying
statistical system. Ruppenier has in particular noticed for the
black holes in general relativity that the scalar curvature
\begin{eqnarray}
R(X) \sim \xi^d,
\end{eqnarray}
where $d$ is the spatial dimension of the statistical system and
the $\xi$ fixes the physical scale \cite{rup2}. The limit $ R(X)
\longrightarrow \infty $ indicates the existence of certain
critical points or phase transitions in the underlying statistical
system.
The fact that ``All the statistical degrees of freedom of a black
hole live on the black hole event horizon'' signifies that the
state-space scalar curvature, as the intrinsic geometric
invariant, indicates an average number of correlated Plank areas
on the event horizon of the black hole \cite{rup22}. In this
concern, Ref.\cite{RuppeinerPRD78} offers interesting physical
properties of the thermodynamic scalar curvature and phase
transitions in Kerr-Newman black holes. Ruppenier has further
conjectured that the global correlations can be expressed by the
following statements: (a) The zero state-space scalar curvature
indicates certain bits of information on the event horizon,
fluctuating independently of each other. (b) The diverging scalar
curvature signals a phase transition indicating highly correlated
pixels of the informations.
\section{Some Physical Motivations}
\subsection{Extremal Black Holes}
The state-space of the extremal black hole configuration is a
reduced space comprising of the states which respect the
extremality (BPS) condition. The state-spaces of the extremal
black holes show an intrinsic geometric description. Our intrinsic
geometric analysis offers a possible zero temperature
characterization of the limiting extremal black brane attractors.
From the gauge/ gravity correspondence, the existence of
state-space geometry could be relevant to the boundary gauge
theories, which have a finitely many countable set of conformal
field theory states.
\subsection{Nonextremal Black Holes}
We shall analyze the state-space geometry of nonextremal black
holes by the addition of anti-brane charge(s) to the entropy of
the corresponding extemal black holes. To interrogate the
stability of a chosen black hole system, we shall investigate the
question that the underlying metric $ g_{ij}(X_i)= - \partial_i
\partial_j S( X_1, X_2,\dots, X_n)$ should provide a
nondegenerate state-space manifold. The exact dependence varies
case to case. In the next section, we shall proceed in our
analysis with an increasing number of the brane charges and
antibrane charges.
\subsection{Chemical Geometry} \label{chemicalgeometry}
The thermodynamic configurations of nonextremal black holes in
string theory with small statistical fluctuations in a
``canonical'' ensemble are stable if the following inequality
holds
\begin{eqnarray}
\Vert \partial_i \partial_j S( X_1, X_2,\dots, X_n) \Vert < 0
\end{eqnarray}
The thermal fluctuations of nonextremal black holes, when
considered in the canonical ensemble, give a closer approximation
to the microcanonical entropy
\begin{eqnarray} \label{microentropy}
S = S_0- \frac{1}{2} \ln(CT^2)+ \cdots
\end{eqnarray}
In the Eq. (\ref{microentropy}), the $S_0$ is the entropy in the
``canonical'' ensemble and $C$ is the specific heat of the black
hole statistical configuration. At low temperature, the quantum
effects dominate and the above expansion does not hold anymore.
The stability condition of the canonical ensemble is just $C > 0$.
In other words, the Hessian function of the internal energy with
respect to the chemical variables, \textit{viz.}, $\{ x_1,
x_2,\dots, x_n\}$, remains positive definite. Hence, the energy as
the function of the $\{ x_1, x_2,\dots, x_n\}$ satisfies the
following condition
\begin{eqnarray}
\Vert \partial_i \partial_j E( x_1, x_2,\dots, x_n) \Vert > 0
\end{eqnarray}
The state-space co-ordinates $\lbrace X^i \rbrace $ and intensive
chemical variables $\lbrace x_i \rbrace$ are conjugate to each
other. In particular, the $\lbrace X^i \rbrace $ are defined as
the Legendre transform of $\lbrace x_i \rbrace$, and thus we have
\begin{equation}
X^i:= \frac{\partial S(x)}{\partial x_i}
\end{equation}
\subsection{Physics of Correlation}
Geometrically, the positivity of the heat capacity $C > 0$ turns
out to be the positivity condition of $g_{ij}>0$, for a given $i$.
In many cases, the state-space stability restriction on the
parameters of the black hole corresponds to the situation away
from the extremality condition, \textit{viz.}, $r_{+} = r_{-}$.
Far from the extremality condition, even at the zero antibrane
charge or angular momentum, we find that there is a finite value
of the thermodynamic scalar curvature, unlike the nonrotating or
only brane charged configurations. It turns out that the
state-space geometry of the two charge extremal configurations is
flat. Thus, the Einstein-Hilbert contributions lead to a non
interacting statistical system. At the tree level, some black hole
configurations turn out to be ill-defined, as well. However, we
anticipate that the corresponding state-space configuration would
become well-defined when a sufficient number of higher derivative
corrections \cite{grqc9307038,grqc9502009,grqc9912119,9305016} is
taken into account with respect to the
$\alpha^{\prime}$-corrections and the string loop $l_s$
corrections. For the BTZ black holes \cite{bnt3}, we notice that
the large entropy limit turns out to be the stability bound,
beyond which the underlying quantum effects dominate.

For the black hole in string theory, the Ricci scalar of the
state-space geometry is anticipated to be positive definite with
finitely many higher order corrections. For nonextremal black
brane configurations, which are far from the extremality
condition, such effects have been seen from the nature of the
state-space scalar curvature $R(S( X_1, X_2,\dots, X_n))$. Indeed,
Refs.\cite{bnt2,bnt5} indicate that the limiting state-space
scalar curvature $R(S( X_1, X_2,\dots, X_n))\vert_{no \
antichagre} \neq 0$ gives a set of stability bounds on the
statistical parameters. Thus, our consideration yields a
classification of the domain of the parameters and global
correlation of a nonextremal black hole.

\subsection{String Theory Perspective} \label{stringtheory}
In this subsection, we recall a brief notion of entropy of a
general string theory black brane configuration from the viewpoint
of the counting of the black hole microstates \cite{9601029v2,
9512078v1,9601029v2,MSW,9602043v2,9603109v1,9603195v1,9603061v2,ref1}.
Given a string theory configuration, the choice of compactification 
\cite{Witten} chosen is the factorization of the type $\mathcal M_{(3,1)}\times M_6$, 
where $M_6$ is a compact internal manifold. From the perspective of statistical 
ensemble theory, we shall express the entropy of a non-extremal black hole as
the function of the numbers of branes and antibranes. Namely, for the charged 
black holes, the electric and magnetic charges $(q_i,p_i)$ form a
coordinate chart on the state-space manifold. In this case, for a
given ensemble of $D$-branes, the coordinate $q_i$ is defined as
the number of the electric branes and $p_i$ as the number of the
magnetic branes. Towards the end of this paper, we shall offer
further motivation for the consideration of the state-space
geometry of large charged non-spherical horizon black holes in
spacetime dimensions $D \ge 5$. In this concern, the Ref.
\cite{ref1} plays a central role towards the formation of the
lower dimensional black hole configuration. Namely, for the torus
compatifications, the exotic branes play an important role
concerning the physical properties of supertubes, the $D_0$-$F_1$
system and associated counting of the black hole microstates.

In what follows, we consider the four dimensional string theory
black holes in a given duality basis of the charges $(q_i,p_i)$.
From the perspective of string theory, the exotic branes and
non-geometric configurations offer interesting fronts for the
black holes in three spacetime dimensions. In general, such
configurations could carry a dipole or a higher pole charge, and
they leave the four dimension black hole configuration
asymptotically flat. In fact, for the spacetime dimensions $D \ge
4$, Ref.  \cite{ref1} shows that a charge particle corresponds to
an underlying gauge field, modulo $U$-duality transformations.
From the perspective of non-extremal black holes, by taking
appropriate boundary condition, namely, the unit asymptotic limit
of the harmonic function which defines the spacetime metric, one
can chose the spacetime regions such that the supertube effects
arising from non-exotic branes can effectively be put off in an
asymptotically flat space \cite{ref1}.
This allows one to compute the Arnowitt-Deser-Misner (ADM) mass
of the asymptotic black hole. From the viewpoint of the statistical
investigation, the dependence of the mass to the entropy of a
non-extremal black hole comes from the contribution of the
antibranes to the counting degeneracy of the states.
\section{Extremal Black Holes in String Theory}
\subsection{Two Charge Configurations}
The state-space geometry of the two charge extremal configurations
is analyzed in terms of the winding modes and the momentum modes
of an excited string carrying $ n_1 $ winding modes and $ n_p $
momentum modes. In the large charge limit, the microscopic entropy
obtained by the degeneracy of the underlying conformal field
theory states reduces to the following expression
\begin{equation}
S_{micro}= 2\sqrt{2n_1 n_p}
\end{equation}
The microscopic counting can be accomplished by considering an
ensemble of weakly interacting D-branes \cite{9512078v1}. The
counting entropy and the macroscopic attractor entropy of the two
charged black holes in string theory which have a $n_4$ number of
$D_4$ branes and a $n_0$ number of $D_0$ branes match and thus we
have
\begin{equation}
S_{micro}= 2 \pi \sqrt{n_0 n_4}= S_{macro}
\end{equation}
In this case, the components of underlying state-space metric
tensor are
\begin{eqnarray}
g_{n_0n_0}= \frac{\pi}{2 n_0} \sqrt{\frac{n_4}{n_0}}, \ 
g_{n_0n_4}= -\frac{\pi}{2} \frac{1}{\sqrt{n_0n_4}}, \ 
g_{n_4n_4}= \frac{\pi}{2 n_4} \sqrt{\frac{n_0}{n_4}}
\end{eqnarray}
The diagonal pair correlation functions remain positive definite
\begin{eqnarray}
g_{n_in_i}> 0 \ \forall \ i \in \{0,4 \} \mid n_i > 0 
g_{n_4n_4}> 0 \ \forall \ (n_0,n_4)
\end{eqnarray}
For distinct $i,j \in \lbrace 0,4 \rbrace $, the state-space pair
correlation functions admit
\begin{eqnarray}
\frac{g_{ii}}{g_{jj}}= (\frac{n_j}{n_i})^2, \ 
\frac{g_{ij}}{g_{ii}}= -\frac{n_i}{n_j}
\end{eqnarray}
The global properties of fluctuating two charge $D_0$-$D_4$
extremal configurations are determined by possible principle
minors. The first minor constraint $p_1>0$ directly follows from
the positivity of the first component of metric tensor
\begin{eqnarray}
p_1&=& \frac{\pi}{2 n_0} \sqrt{ \frac{n_4}{n_0}}
\end{eqnarray}
The determinant of the metric tensor $p_2:= g(n_0,n_4)$ vanishes
identically for all allowed values of the parameters Thus, the
leading order large charge extremal black branes having (i) a
$n_0$ number of $D_0$-branes and a $n_4$ number of $D_4$; or (ii)
excited strings with a $n_1$ number of windings and a $n_p$ number
of momenta, where either set of charges form local coordinates on
the state-space manifold, find degenerate intrinsic state-space
configurations. For a given configuration entropy $S_0:=2\pi c$,
the constant entropy curve can be depicted as the rectangular
hyperbola
\begin{equation}
n_0 n_4= c^2
\end{equation}
The intrinsic state-space configuration depends on the attractor
values of the scalar fields which arise from the chosen string
compactification. Thus, the possible state-space Ruppenier
geometry may become well defined against further higher derivative
$\alpha^{\prime}$-corrections. In particular, the determinant of
the state-space metric tensor may take positive/ negative definite
values over the domain of brane charges. We shall illustrate this
point in a bit more detail in the subsequent consideration with a
higher number of charges and anticharges.
\subsection{Three Charge Configurations}
From the consideration of the two derivative Einstein-Hilbert
action, the Ref.\cite{9601029v2} shows that the leading order
entropy of the three charge $D_1$-$D_5$-$P$ extremal black holes
is
\begin{eqnarray} \label{entropy3e}
S_{micro}= 2 \pi \sqrt{n_1 n_5 n_p}= S_{macro}
\end{eqnarray}
The concerned components of state-space metric tensor are given in the Appendix(A).
Hereby, it follows further that the local state-space metric constraints
are satisfied as
\begin{eqnarray}
g_{n_in_i}&>& 0 \ \forall \ i \in \{1,5,p\} \mid n_i > 0
\end{eqnarray}

For distinct $i,j \in \lbrace 1,5 \rbrace $ and $p$, the list of
relative correlation functions is dipicted in the Appendix(A).
Further, we see that the local stabilities pertaining to the 
lines and two dimensional surfaces of the state-space manifold 
are measured as
\begin{eqnarray}
p_1= \frac{\pi}{2n_1} \sqrt{\frac{n_5 n_p}{n_1}}, \ 
p_2= -\frac{\pi^2}{4n_1n_5^2n_p} (n_p^2 n_1+ n_5^3)
\end{eqnarray}
The stability of the entire equilibrium phase-space configurations
of the $D_1$-$D_5$-$P$ extremal black holes is determined by the
$p_3:=g$ determinant of the state-space metric tensor
\begin{eqnarray}
\Vert g \Vert= -\frac{1}{2}\pi^3 (n_1 n_5 n_p)^{-1/2}
\end{eqnarray}
The universal nature of statistical interactions and the other
properties concerning Maldacena, Strominger and Witten (MSW)
rotating black branes \cite{MSW} are elucidated by the state-space
scalar curvature
\begin{eqnarray}
R(n_1,n_5,n_p)= \frac{3}{4 \pi \sqrt{n_1 n_5 n_p}}
\end{eqnarray}
The constant entropy (or scalar curvature) curve defining the
state-space manifold is the higher dimensional hyperbola
\begin{eqnarray}
n_1 n_5 n_p= c^2,
\end{eqnarray}
where $c$ takes respective values of $(c_S,c_R)=(S_0/ 2 \pi, 3/ 4
\pi R_0)$. In Refs.\cite{bnt2,bnt5,bnt8,bnt9}, we have shown that
similar results hold for the state-space configuration of the four
charge extremal black holes.
\section{Nonextremal Black Holes in String Theory}
\subsection{Four Charge Configurations}
The state-space configuration of the nonextremal $D_1$-$D_5$ black
holes is considered with nonzero momenta along the clockwise and
anticlockwise directions of the Kaluza-Klein compactification
circle $ S^1$. Following Ref.\cite{9602043v2}, the microscopic
entropy and the macroscopic entropy match for given total mass and
brane charges
\begin{eqnarray} \label{entropy4ne}
S_{micro}= 2 \pi \sqrt{n_1 n_5} (\sqrt{n_p}+ \sqrt{
\overline{n_p}})= S_{macro}
\end{eqnarray}
The state-space covariant metric tensor is defined as a negative
Hessian matrix of the entropy with respect to the number of $D_1$,
$D_5$ branes $\{ n_i \mid i= 1, 5 \} $ and clockwise-anticlockwise
Kaluza-Klein momentum charges $\{n_p, \overline{n_p} \}$.
Herewith, we find that the components of the metric tensor take
elagent fomrs. The corresponding expressions are given in the 
Appendix(B).
As in the case of the extremal configurations, the state-space
metric satisfies the following constraints
\begin{eqnarray}
g_{n_in_i}> 0 \ \forall \ i= 1,5; \ 
g_{n_an_a}> 0 \ \forall \ a= p, \overline{p}
\end{eqnarray}

Furthermore, the scaling relations for distinct $i,j \in \lbrace 1,5 \rbrace $ and $p$, 
concerning the list of relative correlation functions is offered in the Appendix(B).
In this case, we find that the stability criteria of the possible surfaces and hyper-surfaces
of the underlying state-space configuration are determined by the
positivity of the following principle minors
\begin{eqnarray}
p_0&=& 1 , \ 
p_1= \frac{\pi}{2}\sqrt{\frac{n_5}{n_1^{3}}}(\sqrt{n_p}+{\sqrt{\overline{n_p}}}) \nonumber \\
p_2&=& 0, \ 
p_3=
-\frac{1}{2n_p}\frac{\pi^3}{\sqrt{n_1n_5}}(\sqrt{n_p}+\sqrt{\overline{n_p}})
\end{eqnarray}
The complete local stability of the full nonextremal $D_1$-$D_5$
black brane state-space configuration is ascertained by the
positivity of the determinant of the metric tensor
\begin{eqnarray}
g(n_1, n_5, n_p,\overline{n_p})=
-\frac{1}{4}\frac{\pi^4}{(n_p\overline{n_p})^{3/2}}(\sqrt{n_p}+\sqrt{\overline{n_p}})^2
\end{eqnarray}
The global state-space properties concerning the four charge
nonextremal $D_1$-$D_5$ black holes are determined by the
regularity of the invariant scalar curvature
\begin{eqnarray} \label{rnpnpbar}
R(n_1, n_5, n_p, \overline{n_p})= \frac{9}{4 \pi \sqrt{n_1 n_5}}
(\sqrt{n_p}+ {\sqrt{\overline{n_p}}})^{-6} f(n_p, \overline{n_p}),
\end{eqnarray}
where the function $f(n_p, \overline{n_p})$ of two momenta $ (n_p,
\overline{n_p}) $ running in opposite directions of the
Kaluza-Klein circle $S^1$ has been defined as
\begin{eqnarray} \label{fnpnpbar}
f(n_p, \overline{n_p}):= n_p^{5/2}+ 10 n_p^{3/2} \overline{n_p}+ 5
n_p^{1/2}\overline{n_p}^{2}+ 5 n_p^{2}\overline{n_p}^{1/2}+ 10 n_p
\overline{n_p}^{3/2}+ \overline{n_p}^{5/2}
\end{eqnarray}
By noticing the Pascal coefficient structure in the Eqn.(\ref{fnpnpbar}), 
we see that the above function $f(n_p, \overline{n_p})$ can be factorized as 
\begin{eqnarray} 
f(n_p, \overline{n_p})=(n_p + \overline{n_p})^5
\end{eqnarray}
Thus, Eqn.(\ref{rnpnpbar}) leads to the following state-space scalar curvature
\begin{eqnarray} 
R(n_1, n_5, n_p, \overline{n_p})= \frac{9}{4 \pi \sqrt{n_1 n_5}} \times
\bigg(\frac{1}{\sqrt{n_p}+ {\sqrt{\overline{n_p}}}} \bigg)
\end{eqnarray}
In the large charge limit, the nonextremal $D_1$-$D_5$ black
branes have a nonvanishing small scalar curvature function on the
state-space manifold $(M_4,g)$. This implies an almost everywhere
weakly interacting statistical basis. In this case, the constant
entropy hypersurface is defined by the curve
\begin{eqnarray}
\frac{c^2}{n_1 n_5}= (\sqrt{n_p}+ \sqrt{\overline{n_p}})^{2}
\end{eqnarray}
As in the case of two charge $D_0$-$D_4$ extremal black holes and
$D_1$-$D_5$-$P$ extremal black holes, the constant $c$ takes the
same value of $c:=S_0^2/ 4 \pi^2$. For a given state-space scalar
curvature $k$, the constant state-space curvature curves take the
following form
\begin{eqnarray}
f(n_p, \overline{n_p})= k \sqrt{n_1 n_5}
(\sqrt{n_p}+{\sqrt{\overline{n_p}}})^{6}
\end{eqnarray}
\subsection{Six Charge Configurations}
We now extrapolate the state-space geometry of four charge
nonextremal $D_1$-$D_5$ solutions for nonlarge charges, where we
are no longer close to an ensemble of supersymmetric states. In
Ref.\cite{9603109v1}, the computation of the entropy of all such
special extremal and near-extremal black hole configurations has
been considered. The leading order entropy as a function of
charges $\lbrace n_i \rbrace$ and anticharges $\lbrace m_i
\rbrace$ is
\begin{eqnarray} \label{entropy6ne}
S(n_1,m_1,n_2,m_2,n_3,m_3):= 2 \pi (\sqrt{n_1} + \sqrt{m_1})
(\sqrt{ n_2 }+ \sqrt{m_2}) (\sqrt{n_3}+ \sqrt{m_3})
\end{eqnarray}
For given charges $i, j \in A_1:= \{n_1,m_1\}$; $k, l \in A_2:=
\{n_2,m_2\}$; and $m,n \in A_3:= \{n_3,m_3\}$, the intrinsic
state-space pair correlations are in precise accordance with the
underlying macroscopic attractor configurations which are being
disclosed in the special leading order limit of the nonextremal
$D_1$-$D_5$ solutions. The components of the covariant state-space
metric tensor over generic nonlarge charge domains are not difficult
to compute, and indeed, we have offered their corresponding expressions 
in the Appendix(C).

For all finite $(n_i, m_i), i=1,2,3$, the components involving
brane-brane state-space correlations $g_{n_in_i}$ and
antibrane-antibrane state-space correlations $g_{m_im_i}$ satisfy
the following positivity conditions
\begin{eqnarray}
g_{n_in_i}> 0, \ \ 
g_{m_im_i}> 0
\end{eqnarray}
The distinct $\{ n_i, m_i \mid i \in \lbrace 1,2,3 \rbrace \}$
describing six charge string theory black holes have three types
of relative pair correlation functions. The corresponging expressions
of the relative statistical correlation functions are given in the 
Appendix(C).

Notice hereby that the scaling relations remain similar to those obtained 
in the previous case, except (i) the number of relative correlation
functions has been increased, and (ii) the set of cross ratios,
\textit{viz.}, $\{\frac{g_{ij}}{g_{kl}}, \frac{g_{kl}}{g_{mn}},
\frac{g_{ij}}{g_{mn}}\}$ being zero in the previous case, becomes
ill-defined for the six charge state-space configuration.
Inspecting the specific pair of distinct charge sets $A_i$ and
$A_j$, there are now 24 types of nontrivial relative correlation
functions. The set of principle components denominator ratios
computed from the above state-space metric tensor reduces to
\begin{eqnarray}
\frac{g_{ij}}{g_{kk}}&=& 0 \ \forall \ i, \ j, \ k \in \{n_1, m_1,
n_2, m_2, n_3, m_3\}
\end{eqnarray}
For given $i, j \in A_1:= \{n_1,m_1\}$; $k, l \in A_2:=
\{n_2,m_2\}\}$; $m,n \in A_3:= \{n_3,m_3\}$, and $g_{n_i m_i}=0$,
there are the total 15 types of trivial relative correlation
functions. There are five such trivial ratios in each family $\{
A_i \mid \ i= 1,2,3\}$. The local stability of the higher charged
string theory nonextremal black holes is given by
\begin{eqnarray}
p_1 &=& \frac{\pi}{2n_1^{3/2}} ( \sqrt{n_2} + \sqrt{m_2} )( \sqrt{n_3}+ \sqrt{m_3} ) \nonumber \\
p_2 &=& \frac{1}{4} \frac{\pi^2}{(n_1m_1)^{3/2}}(\sqrt{n_2}+\sqrt{m_2})^2(\sqrt{n_3}+\sqrt{m_3})^2 \nonumber \\
p_3 &=&  \frac{1}{8} \frac{\pi^3}{(n_1m_1n_2)^{3/2}} \sqrt{m_2}
(\sqrt{n_3}+\sqrt{m_3})^3 (\sqrt{n_2}+\sqrt{m_2})
(\sqrt{n_1}+\sqrt{m_1}) \nonumber \\
p_4 &=& 0 
\end{eqnarray}
The principle minor $p_5$ remains nonvanishing for all values of
charges on the constituent brane and anti branes. In general, by an explicit
calculation, we find that the hyper-surface minor $p_5$ takes the following
nontrivial value  
\begin{eqnarray}
p_5= -\frac{1}{8} \frac{\pi^5}{(n_1m_1n_2m_2)^{3/2}n_3}
(\sqrt{n_1} + \sqrt{m_1})^3 (\sqrt{ n_2 }+ \sqrt{m_2})^3 (\sqrt{n_3}+ \sqrt{m_3})^3
\end{eqnarray}
Specifically, for an identical value of
the brane and antibrane charges, the minor $p_5$ reduces to
\begin{eqnarray}
p_5(k)= -64 \frac{\pi^5}{k^{5/2}}
\end{eqnarray}
The global stability on the full state-space configuration is
carried forward by computing the determinant of the metric tensor
\begin{eqnarray}
\Vert g \Vert&=& 
-\frac{1}{16} \frac{\pi^6}{(n_1m_1n_2m_2n_3m_3)^{3/2}}
(\sqrt{n_1} + \sqrt{m_1})^4 \nonumber \\&&  
(\sqrt{ n_2 }+ \sqrt{m_2})^4 (\sqrt{n_3}+ \sqrt{m_3})^4
\end{eqnarray}
The underlying state-space configuration remains nondegenerate for
the domain of given nonzero brane antibrane charges, except for
extreme values of the brane and antibrane charges $\{ n_i, m_i
\}$, when they belong to the set
\begin{eqnarray}
B&:=&\lbrace \ (n_1,n_2,n_3,m_1,m_2,m_3) \mid  
(n_i,m_i) = (0,0), (\infty, \infty),\ some\ i 
\rbrace
\end{eqnarray}
among the given brane-antibrane pairs $\{(n_1,m_1), (n_2,m_2), (n_3,m_3)\}$.   
The component $ R_{n_1 n_2 m_3 m_4} $ diverges at the roots of the
two variables polynomials defined as the functions of brane and
antibrane charges
\begin{eqnarray}
f_1(n_2,m_2)&=& n_2^4 m_2^3+ 2 (n_2 m_2)^{7/2} + n_2^3 m_2^4 \nonumber \\
f_2(n_3,m_3)&=& m_3^{9/2} n_3^4 + n_3^4 m_3^{9/2}
\end{eqnarray}
However, the component $ R_{n_3,m_3,n_3,m_3} $ with an equal
number of brane and anti-brane charges diverges at a root of a
single higher degree polynomial
\begin{eqnarray}
f(n_1,m_1,n_2,m_2,n_3,m_3)&:=& n_2^4 m_2^3 n_3^{9/ 2} m_3^4 +
n_2^4 m_2^3 n_3^4 m_3^{9/2}+ 2 n_2^{7/2} m_2^{7/2} n_3^{9/2}
m_3^4+\nonumber \\&& 2 n_2^{7/2} m_2^{7/2}n_3^4 m_3^{9/2}+ n_2^3
m_2^4 n_3^{9/2} m_3^4+ n_2^3 m_2^4 n_3^4 m_3^{9/2}
\end{eqnarray}
Herewith, from the perspective of state-space global invariants,
we focus towards for the limiting nature of the underlying ensemble. 
Thus, we may chose the equal charge and anticharge limit by defining $m_i:=k$ and
$n_i:=k$ for the calculation of the Ricci scalar. In this case, 
we find the following small negative curvature scalar
\begin{eqnarray}
R(k)= -\frac{15}{16} \frac{1}{\pi k^{3/2}}
\end{eqnarray}
Furrther, the physical meaning of taking an equal value of the charges and anticharges lies in the ensemble
theory, \textit{viz.} in the thermodynamic limit, all the statistical fluctuations of the charges and anticharges
approach to a limiting Gaussian fluctuations. In this sense, we can take the average over the concerned 
individual Gaussian fluctuations. This shows that the limiting statistical ensemble of nonextremal nonlarge charge $D_1$-$D_5$ solutions
yields an attractive state-space configuration. Finally, such a limiting procedure is indeed defined by considering the standard deviations 
of the equal integer charges and anticharges, and thus our interest in calculating the limiting Ricci scalar
in order to know the nature of the long rang interactions underlying in the system.

For a given entropy $S_0$, the constant entropy hypersurface is again
some nonstandard curve
\begin{eqnarray}
(\sqrt{n_1} + \sqrt{m_1}) (\sqrt{ n_2 }+ \sqrt{m_2}) (\sqrt{n_3}+
\sqrt{m_3}) = c,
\end{eqnarray}
where the real constant $ c $ takes the precise value of
$S_0/2\pi$.
\subsection{Eight Charge Configurations}
From the perspective of the higher charged anticharged black hole
configurations in string theory, let us systematically analyze the
underlying statistical structures. In this case, the state-space
configuration of the nonextremal black hole involves finitely many
nontrivially circularly fibered Kaluza-Klein monopoles. In this
process, we enlist the complete set of nontrivial relative
state-space correlation functions of the eight charged anticharged
configurations, with respect to the lower parameter
configurations, as considered in Refs.\cite{bnt2,bnt5}. There have
been calculations of the entropy of the extremal, near-extremal
and general nonextremal solutions in string theory, see for
instances \cite{9603195v1,9603061v2}. Inductively, the most
general charge anticharge nonextremal black hole has the following
entropy \ba \label{entropy8ne} S(n_1,m_1,n_2,m_2,n_3,m_3,n_4,m_4) &=& 2 \pi
\prod_{i=1}^4(\sqrt{n_i}+\sqrt{m_i}). \ea For the distinct $i,j,k
\in \{1,2,3,4\}$, we find that the components of the metric tensor
are \ba g_{n_in_i} &=& \frac{\pi}{2n_i^{3/2}} \prod_{j \ne i}
(\sqrt{n_j}+\sqrt{m_j}), \nn
g_{n_in_j} &=& -\frac{\pi}{2 (n_i n_j)^{1/2}} \prod_{i \ne k \ne
j} (\sqrt{n_k}+\sqrt{m_k}), \nn
g_{n_im_i} &=& 0, \nn
g_{n_im_j} &=& -\frac{\pi}{2 (n_i m_j)^{1/2}} \prod_{i \ne k \ne
j} (\sqrt{n_k}+\sqrt{m_k}), \nn
g_{m_im_i} &=& \frac{\pi}{2 m_i^{3/2}} \prod_{j \ne i}
(\sqrt{n_j}+\sqrt{m_j}),\nn
g_{m_im_j} &=& -\frac{\pi}{2 (m_i m_j)^{1/2}} \prod_{i \ne k \ne
j} (\sqrt{n_k}+\sqrt{m_k}).\ea
From the above depiction, it is evident that the principle
components of the state-space metric tensor $\lbrace g_{n_in_i},
g_{m_im_i} \vert \ i=1, 2, 3, 4 \rbrace$ essentially signify a set
of definite heat capacities (or the related compressibilities)
whose positivity in turn apprises that the black brane solutions
comply with an underlying equilibrium statistical configuration.
For an arbitrary number of the branes $\{n_i\}$ and antibranes
$\{m_i\}$, we find that the associated state-space metric
constraints as the diagonal pair correlation functions remain
positive definite.  In particular, $\forall \ i \in \{1,2,3,4 \}$,
it is clear that we have the following positivity conditions
\begin{eqnarray}
g_{n_in_i}> 0 \ \mid n_i, m_i > 0, \ \ 
g_{m_im_i}> 0 \ \mid n_i, m_i > 0\end{eqnarray}
As observed in Refs.\cite{bnt2,bnt5}, we find that the ratios of
diagonal components vary inversely with a multiple of a
well-defined factor in the underlying parameters, \textit{viz.},
the charges and anticharges, which changes under the Gaussian
fluctuations, whereas the ratios involving off diagonal components
in effect uniquely inversely vary, in the parameters of the chosen
set $A_i$ of equilibrium black brane configurations. This suggests
that the diagonal components weaken in a relatively controlled
fashion into an equilibrium, in contrast with the off diagonal
components, which vary over the domain of associated parameters
defining the $D_1$-$D_5$-$P$-$KK$ nonextremal nonlarge charge
configurations. In short, we can easily substantiate, for the
distinct $x_i:=(n_i, m_i) \mid i \in \lbrace 1,2,3,4 \rbrace $
describing eight (anti)charge string theory black holes, that the
relative pair correlation functions have distinct types of
relative correlation functions. Apart from the zeros, infinities
and similar factorizations, we see that the nontrivial relative
correlation functions satisfy the following scaling relations
\begin{eqnarray}
\frac{g_{x_ix_i}}{g_{x_jx_j}}&=& (\frac{x_j}{x_i})^{3/2} 
\frac{\sqrt{n_j}+\sqrt{m_j}}{\sqrt{n_i}+\sqrt{m_i}}, \nonumber \\
\frac{g_{x_ix_j}}{g_{x_kx_l}}&=& (\frac{x_ix_j}{x_kx_l})^{-1/2}
\frac{\prod_{i \ne p \ne j} ( \sqrt{n_p}+\sqrt{m_p})}{\prod_{k \ne
q \ne l} (\sqrt{n_q}+\sqrt{m_q})}, \nonumber \\
\frac{g_{x_ix_i}}{g_{x_ix_k}}&=&-\sqrt{(\frac{x_k}{x_i^2})}
\frac{\prod_{p \ne i }( \sqrt{n_p}+\sqrt{m_p})}{\prod_{i \ne q \ne
k} (\sqrt{n_q}+\sqrt{m_q})}.
\end{eqnarray}
As noticed in Refs.\cite{bnt2,bnt5}, it is not difficult to
analyze the statistical stability properties of the eight charged
anticharged nonextremal black holes, viz. we can compute the
principle minors associated with the state-space metric tensor and
thereby argue that all the principle minors must be positive
definite, in order to have a globally stable configuration. In the
present case, it turns out that the above black hole is stable
only when some of the charges and or anticharges are held fixed or
take specific values such that $p_i > 0$ for all the dimensions of
the state-space manifold. From the definition of the Hessian
matrix of the associated entropy concerning the most general
nonextremal nonlarge charged black holes, we observe that some of
the  the principle minors $p_i$ are indeed nonpositive. In fact,
we discover an uniform local stability criteria on the three
dimensional hyper-surfaces, two dimensional surface and the one dimensional
line of the underlying state-space manifold. In order to simplify the factors 
of the higher principle, we may hereby collect the powers of each factors 
$(\sqrt{n_i}+\sqrt{m_i})$ appearing in the expression of the entropy. 
With this notation, the Appendix(D) provides the corresponding principle
minors of take the most general nonextremal nonlarge charged anticharged 
black hole in string theory involving finitely many nontrivially 
circularly fibered Kaluza-Klein monopoles.

Notice that the heat capacities, as the diagonal components
$g_{ii}$, surface minor $p_2$, hypersurface minors $p_3$, $p_5$,
$p_6$, $p_7$, and the determinant of the state-space metric
tensor, as the highest principle minor $p_8$ are examined as the
functions of the number of branes $n$ and antibranes $m$. Thus,
they describe the nature of the statistical fluctuations in the
vacuum configuration. The corresponding scalar curvature is
offered for an equal number of branes and antibranes $(n= m)$,
which describes the nature of the long range statistical
fluctuations. As per the above evaluation, we have obtained the
exact expressions for the components of the metric tensor,
principle minors, determinant of the metric tensor and the
underlying scalar curvature of the fluctuating statistical
configuration of the eight parameter black holes in string theory.
Qualitatively, the local and the global correlation properties of
the limiting vacuum configuration can be realized under the
statistical fluctuations. The first seven principle minors
describe the local stability properties, and the last minor
describes the global ensemble stability.

The scalar curvature describes the corresponding phase space
stability of the eight parameter black hole configuration. In
general, there exists an akin higher degree polynomial equation on
which the Ricci scalar curvature becomes null, and exactly on
these points the state-space configuration of the underlying
nonlarge charge nonextremal eight charge black hole system
corresponds a noninteracting statistical system. In this case, the
corresponding state-space manifold $(M_8,g)$ becomes free from the
statistical interaction with a vanishing state-space scalar curvature. 
As in case of the six charge configuration, we find interestingly that 
there exists an attractive configuration for the equal number of branes 
$n:=k$ and antibranes $m:=k$. In the limit of a large $k$, the corresponding
system possesses a small negative value of the state-space scalar
curvature \ba R(k) = -\frac{21}{32} \frac{1}{\pi k^2} \ea
Interestingly, it turns out that the system becomes noninteracting
in the limit of $k \rightarrow \infty$. For the case of the $n= k=
m$, we observe that the corresponding principle minors reduce to
the following constant values \ba \{p_i\}_{i=1}^8 &=& \{4 \pi, 16
\pi^2, 32 \pi^3, 0, -2048 \pi^5, -16384 \pi^6, \nn && -163840 \pi^7,
-1048576 \pi^8 \}.\ea In this case, we find that the limiting
underlying statistical system remains stable when at most three of
the parameters, \textit{viz.}, $\{n_i=k= m_i\}$, are allowed to fluctuate. 
Herewith, we find for the case of $n:=k$ and $m:=k$ that the state-space 
manifold of the eight parameter brane and antibrane configuration is free 
from critical phenomena, except for the roots of the determinant. Thus,
the regular state-space scalar curvature is comprehensively
universal for the nonlarge charge nonextremal black brane
configurations in string theory. In fact, the above perception
turns out to be justified from the typical state-space geometry,
\textit{viz.}, the definition of the metric tensor as the negative
Hessian matrix of the duality invariant expression of the black
brane entropy. In this case, we may nevertheless easily observe,
for a given entropy $S_0$, that the constant entropy hypersurface
is given by the following curve \begin{eqnarray} (\sqrt{n_1} +
\sqrt{m_1}) (\sqrt{ n_2 }+ \sqrt{m_2}) (\sqrt{n_3}+ \sqrt{m_3})
(\sqrt{ n_4 }+ \sqrt{m_4}) = c, \end{eqnarray} where $ c $ is a
real constant taking the precise value of $S_0/2\pi$. Under the
vacuum fluctuations, the present analysis indicates that the
entropy of the eight parameter black brane solution defines a
nondegenerate embedding in the viewpoints of intrinsic state-space
geometry. The above state-space computations determine an
intricate set of statistical properties, \textit{viz.}, pair
correlation functions and correlation volume, which reveal the
possible nature of the associated parameters prescribing an
ensemble of microstates of the dual conformal field theory living
on the boundary of the black brane solution. For any black brane
configuration, the above computation hereby shows that we can
exhibit the state-space geometric acquisitions with an appropriate
comprehension of the required parameters, e.g., the charges and
anticharges $\{ n_i, m_i \}$, which define the coordinate charts.
From the consideration of the state-space geometry, we have
analyzed state-space pair correlation functions and the notion of
stability of the most general nonextremal black hole in string
theory. From the perspective of the intrinsic Riemannian geometry,
we find that the stability of these black branes has been divulged
from the positivity of principle minors of the space-state metric
tensor.

Herewith, we have explicitly extended the state-space analysis for
the four charge and four anticharge nonextremal black branes in
string theory. The present consideration of the eight parameter
black brane configurations, where the underlying leading order
statistical entropy is written as a function of the charges $
\lbrace n_i \rbrace $ and anticharges $ \lbrace m_i \rbrace $,
describes the stability properties under the Gaussian
fluctuations. The present consideration includes all the special
cases of the extremal and near-extremal configurations with a
fewer number of charges and anticharges. In this case, we obtain
the standard pattern of the underlying state-space geometry and
constant entropy curve as that of the lower parameter nonextremal
black holes. The local coordinate of the state-space manifold
involves four charges and four anticharges of the underlying
nonextremal black holes. In fact, the conclusion to be drawn
remains the same, as the underlying state-space geometry remains
well-defined as an intrinsic Riemannian manifold $ N:= M_8
\setminus \tilde{B}$, where $ \tilde{B} $ is the set of roots of
the determinant of the metric tensor. In particular, the
state-space configuration of eight parameter black brane solutions
remains nondegenerate for various domains of nonzero brane
antibrane charges, except for the values, when the brane charges
$\{n_i\}$ and antibrane charges $\{ m_i \}$ belong to the set
\begin{eqnarray} \tilde{B}&:=&\lbrace \
(n_1,n_2,n_3,n_4,m_1,m_2,m_3, m_4) \mid (n_i,m_i) = (0,0), (\infty, \infty)\rbrace
\end{eqnarray}
for a given brane-antibrane pair $i \in \{1, 2, 3, 4\}$.   
Our analysis indicates that the leading order statistical behavior
of the black brane configurations in string theory remains intact
under the inclusion of the Kaluza-Klein monopoles. In short, we
have considered the eight charged anticharged string theory black
brane configuration and analyzed the state-space pair correlation
functions, relative scaling relations, stability conditions and
the corresponding global properties. Given a general nonextremal
black brane configuration, we have exposed (i) for what conditions
the considered black hole configuration is stable, (ii) how its
state-space correlations scale in terms of the numbers of branes
and antibranes.
\section{Conclusion and Outlook}
The Ruppenier geometry of two charge leading order extremal black
holes remains flat or ill-defined. Thus, the statistical systems
are respectively noninteracting or require higher derivative
corrections. Whilst, an addition of the third brane charge and
other brane and antibrane charges indicates an interacting
statistical system. The statistical fluctuations in the canonical
ensemble leads to an interacting statistical system, as the scalar
curvature of the state-space takes a nonzero value. We have
explored the state-space geometric description of the charged
extremal and associated charged, anticharged nonextremal black
holes in string theory.

Our analysis illustrates that the stability properties of the
specific state-space hypersurface may exactly be exploited in
general. The definite behavior of the state-space properties, as
accounted in the specific cases suggests that the underlying
hypersurfaces of the state-space configuration include the
intriguing mathematical feature. Namely, we find well defined
stability properties for the generic extremal and nonextremal
black brane configurations, except for some specific values of the
charges and anticharges. With and without the large charge limit,
we have provided explicit forms of the higher principle minors of
the state-space metric tensor for various charged, anticharged,
extremal and nonextremal black holes in string theory. In this
concern, the state-space configurations of the string theory black
holes are generically well-defined and indicate an interacting
statistical basis. Interestingly, we discover the state-space
geometric nature of all possible general black brane
configurations. From the very definition of the intrinsic metric
tensor, the present analysis offers a definite stability character
of string theory vacua.

Significantly, we notice that the related principle minors and the
invariant state-space scalar curvature classify the underlying
statistical fluctuations. The scalar curvature of a class of
extremal black holes and the corresponding nonextremal black
branes is everywhere regular with and without the stringy
$\alpha^{\prime}$-corrections. A nonzero value of the state-space
scalar curvature indicates an interacting underlying statistical
system. We find that the antibrane corrections modify the
state-space curvature, but do not induce phase transitions. In the
limit of an extremal black hole, we construct the intrinsic
geometric realization of a possible thermodynamic description at
the zero temperature.

Importantly, the notion of the state-space of the considered black
hole follows from the corresponding Wald and Cardy entropies. The
microscopic and macroscopic entropies match in the large charge
limit. From the perspective of statistical fluctuations, we
anticipate the intrinsic geometric realization of two point local
correlation functions and the corresponding global correlation
length of the underlying conformal field theory configurations. In
relation to the gauge-gravity correspondence and extremal black
holes, our analysis describes state-space geometric properties of
the corresponding boundary gauge theory.
\subsection*{General Remarks} For distinct $\{ i,j \}$, the
state-space pair correlations of an extremal configurations scale
as
\begin{eqnarray}
\frac{g_{ii}}{g_{jj}}= (\frac{X_j}{X_i})^2, \ 
\frac{g_{ij}}{g_{ii}}= -\frac{X_i}{X_j}
\end{eqnarray}
In general, the black brane configurations in string theory can be
categorized as per their state-space invariants. The underlying
sub-configurations turn out to be well-defined over possible
domains, whenever there exist a respective set of nonzero
state-space principle minors. The underlying full configuration
turns out to be everywhere well-defined, whenever there exists a
nonzero state-space determinant. The underlying configuration
corresponds to an interacting statistical system, whenever there
exists a nonzero state-space scalar curvature. The intrinsic
state-space manifold of  extremal/ non-extremal and supersymmetric/
nonsupersymmetric string theory black holes may intrinsically be
described by an embedding
\begin{equation}
(M_{(n)},g) \hookrightarrow (M_{(n+1)},\tilde{g})
\end{equation}
The extremal state-space configuration may be examined as a restriction
to the full counting entropy with an intrinsic state-space metric tensor
$g \mapsto \tilde{g}\vert_{r_{+}=r_{-}}$. Furthermore, the state-space
configurations of the supersymmetric black holes may be examined as the
BPS restriction of the full space of the counting entropy with an understanding
that the intrinsic state-space metric tensor is defined as $g:=\tilde{g}\vert_{M=M_0}$.
From the perspective of string theory, the restrictions $r_{+}=r_{-}$ and
$M=M_0(P_i,Q_i)$ should be understood as the fact that it has been applied to an
assigned entropy of the non-extremal/ nonsupersymmetric (or nearly extremal/ nearly
supersymmetric) black brane configuration.
This allows one to compute the fluctuations in ADM mass of the black hole.
In the viewpoint of the present research on the state-space geometry, 
it is worth mentioning that the dependence of the mass to the entropy 
of a non-extremal black hole comes from the contribution of the antibranes,
see for instance section \ref{stringtheory}, and so we may examine the
corresponging Weinhold chemical geometry, as mentioned in section 
\ref{chemicalgeometry}.
\subsection*{Future Directions and Open Issues} The state-space
instabilities and their relation to the dual microscopic conformal 
field theories could open up a number of new realizations.
The state-space perspective includes following issues.
\begin{itemize}
 \item Multi-center Gibbons-Hawking solutions \cite{07052564v1,0702146v2}
with generalized base space manifolds having a mixing of positive and negative residues,
see \cite{0604110, 07063786V2} for a perspective development of state-space geometry
by invoking the role of foaming of black holes and plumbing the Abyss
for the microstates counting of black rings.
 \item Dual conformal field theories and string duality
symmetries, see \cite{0412322} for a quantum mechanical perspective
of superconformal black holes and \cite{hepth0508023,hepth0107119}
for the origin of gravitational thermodynamics and the role of 
giant gravitons in conformal field theory.
 \item Stabilization against local and/ or global perturbations, see 
\cite{GRG,HoMy,EEV,GL,LoTeAr,odias} for black brane dynamics, stability 
and critical phenomena. Thus, the consideration of state-space geometry
is well suited for examining the domain of instability. This includes 
Gregory-Laflamme (GL) modes, chemical potential fluctuations, 
electric-magnetic charges and dipole charges, rotational fluctuations 
and the thermodynamic temperature fluctuations for the near-extemal 
and nonextremal black brane solutions. We leave this perspective
of the state-space geometry open for a future research.
\end{itemize}
In general, various $D$ dimensional black brane configurations, 
see for instance \cite{GRG,HoMy,EEV,GL,LoTeAr,odias} for black rings 
in $D>5$ spacetime dimensions with $S^1\times S^{D-3}$ horizon topology, 
and the higher horizon topologies, e.g., $S^1\times S^1 \times S^2$, $S^3\times S^3$, 
etc. offer a platform to extend the consideration of the state-space geometry.

On the other hand, the bubbling black brane solutions,
\textit{viz.}, Lin, Lunin and Maldacena (LLM) geometries
\cite{LLM} are interesting from the perspective of Mathur's
Fuzzball conjecture(s). Form the perspective of the generalized
hyper-K\"ahler manifolds, Mathur's conjecture \cite{0109154v1,0202072v2,0706.3884v1,0804.0552}
reduces to classifying and counting asymptotically flat four dimensional
hyper K\"ahler manifolds \cite{0604110} which have moduli regions of uniform
signature $(+,+,+,+)$ and $(-,-,-,-)$.

Finally, the new physics at the length of the Planck scale
anticipates an analysis of the state-space configurations. In
particular, it materializes that the state-space geometry may be
explored with the parameters of the foam geometries \cite{0604110}, 
and the corresponding empty space virtual black holes, see \cite{LLM}
for the notion of bubbling AdS space and 1/2 BPS geometries. In such cases, 
the local and global statistical correlations, among the parameters of
the microstates of the black hole conformal field theory \cite{maldacena,0412322}, 
would involve the foams of two-spheres. From the perspective of the
string theory, the present exploration thus opens up an avenue for
learning new insights into the promising structures of the black
brane space-time configurations at very small scales.

\section*{Acknowledgements}
This work has been supported in part by the European Research
Council grant n.~226455, \textit{``SUPERSYMMETRY, QUANTUM GRAVITY
AND GAUGE FIELDS (SUPERFIELDS)"}.

B.N.T. would like to thank \textit{Prof. V. Ravishankar} for his
support and encouragements towards the research in string theory.
This work was conducted during the period B.N.T. served as a
postdoctoral research fellow at the \textit{INFN-Laboratori
Nazionali di Frascati, Roma, Italy}.

\appendix
\section*{Appendix}
In this appendix, we provide explicit forms of the state-space 
correlation arising from the metric tensor of the charged (non)extremal 
(non)large black holes in string theory. In fact, our analysis illustrates
that the stability properties of the specific state-space hypersurface 
may exactly be exploited in general. The definite behavior of state-space 
properties, as accounted in the concerned main sections suggests that the 
various intriguing hypersurfaces of the state-space configuration include 
the nice feature that they do have definite stability properties, 
except for some specific values of the charges and anticharges. 

As mentioned in the main sections, these configurations are generically 
well-defined and indicate an interacting statistical basis. Herewith, 
we discover that the state-space geometry of the general black brane
configurations in string theory indicate the possible nature of
the underlying statistical fluctuations. Significantly, we notice
from the very definition of the intrinsic metric tensor that the
related the statistical pair correlation functions and relative 
statistical correlation functions take the following eaxct expressions

\section{Correlations for Three Charge Configurations}
Following the notion of the fluctuations, we see from the 
Hessian of the entropy Eqn.(\ref{entropy3e}) that the
components of state-space metric tensor are
\begin{eqnarray}
g_{n_1n_1}&=& \frac{\pi}{2n_1} \sqrt{\frac{n_5 n_p}{n_1}}, \ 
g_{n_1n_5}= -\frac{\pi}{2} \sqrt{\frac{n_p}{n_1 n_5}} \nonumber \\
g_{n_1n_p}&=& -\frac{\pi}{2} \sqrt{\frac{n_5}{n_1 n_p}}, \ 
g_{n_5n_5}= \frac{\pi}{2n_5} \sqrt{\frac{n_1 n_p}{n_5}} \nonumber \\
g_{n_5n_p}&=& -\frac{\pi}{2} \sqrt{\frac{n_1}{n_5 n_p}}, \ 
g_{n_pn_p}= \frac{\pi}{2n_p} \sqrt{\frac{n_1 n_5}{n_p}}
\end{eqnarray}

For distinct $i,j \in \lbrace 1,5 \rbrace $ and $p$, the list of
relative correlation functions follows the scalings
\begin{eqnarray}
\frac{g_{ii}}{g_{jj}}&=& (\frac{n_j}{n_i})^2, \ 
\frac{g_{ii}}{g_{pp}}= (\frac{n_p}{n_i})^2, \ 
\frac{g_{ii}}{g_{ij}}= -(\frac{n_j}{n_i}) \nonumber \\
\frac{g_{ii}}{g_{ip}}&=& -(\frac{n_p}{n_i}), \ 
\frac{g_{ip}}{g_{jp}}= (\frac{n_j}{n_i}), \ 
\frac{g_{ii}}{g_{jp}}= -(\frac{n_jn_p}{n_i^2})\nonumber \\
\frac{g_{ip}}{g_{pp}}&=& -(\frac{n_p}{n_i}), \ 
\frac{g_{ij}}{g_{ip}}= (\frac{n_p}{n_j}), \ 
\frac{g_{ij}}{g_{pp}}= -(\frac{n_p^2}{n_in_j})
\end{eqnarray}

\section{Correlations for Four Charge Configurations}

For the given entropy as in Eqn.(\ref{entropy4ne}),
we find that the components of the metric tensor are
\begin{eqnarray}
g_{n_1 n_1}&=&  \frac{\pi}{2}\sqrt{\frac{n_5}{n_1^{3}}}(\sqrt{n_p}+{\sqrt{\overline{n_p}}}), \ 
g_{n_1 n_5}= -\frac{\pi}{2\sqrt{n_1 n_5}} (\sqrt{n_p}+{\sqrt{\overline{n_p}}})\nonumber \\
g_{n_1 n_p}&=& -\frac{\pi}{2} \sqrt{\frac{n_5}{n_1 n_p}}, \ 
g_{n_1 \overline{n_p}}= -\frac{\pi}{2} \sqrt{\frac{n_5}{n_1 \overline{n_p}}} \nonumber \\
g_{n_5 n_5}&=&  \frac{\pi}{2}\sqrt{\frac{n_1}{n_5^{3}}}(\sqrt{n_p}+{\sqrt{\overline{n_p}}}), \ 
g_{n_5 n_p}= -\frac{\pi}{2} \sqrt{\frac{n_1}{n_5 n_p}} \nonumber \\
g_{n_5 \overline{n_p}}&=& -\frac{\pi}{2} \sqrt{\frac{n_1}{n_5 \overline{n_p}}}, \ 
g_{n_p n_p}= \frac{\pi}{2} \sqrt{\frac{n_1 n_5}{n_p^{3}}} \nonumber \\
g_{n_p \overline{n_p}}&=& 0, \ 
g_{\overline{n_p} \overline{n_p}}= \frac{\pi}{2} \sqrt{\frac{n_1
n_5}{\overline{n_p}^{3}}}
\end{eqnarray}

For distinct $i, j \in \{1,5\}$, and $k, l \in \{p,\overline{p}\}$
describing four charge nonextremal $D_1$-$D_5$-$P$-$\overline{P}$
black holes, the statistical pair correlations consist  of the
following scaling relations
\begin{eqnarray}
\frac{g_{ii}}{g_{jj}}&=& (\frac{n_j}{n_i})^2, \ 
\frac{g_{ii}}{g_{kk}}= \frac{n_k}{n_i^2} \sqrt{n_k}(\sqrt{n_p}+\sqrt{ \overline{n_p}}), \ 
\frac{g_{ii}}{g_{ij}}= -\frac{n_j}{n_i} \nonumber \\
\frac{g_{ii}}{g_{ik}}&=& -\frac{\sqrt{n_k}}{n_i}(\sqrt{n_p}+\sqrt{ \overline{n_p}}), \ 
\frac{g_{ik}}{g_{jk}}= \frac{n_j}{n_i}, \ 
\frac{g_{ii}}{g_{jk}}= -\frac{n_j}{n_i^2} \sqrt{n_k}(\sqrt{n_p}+\sqrt{ \overline{n_p}}) \nonumber \\
\frac{g_{ik}}{g_{kk}}&=& -\frac{n_k}{n_i}, \ 
\frac{g_{ij}}{g_{ik}}= \frac{\sqrt{n_k}}{n_j} (\sqrt{n_p}+\sqrt{ \overline{n_p}}), \ 
\frac{g_{ij}}{g_{kk}}= - \frac{n_k}{n_in_j} \sqrt{n_k}(\sqrt{n_p}+\sqrt{ \overline{n_p}}) \nonumber \\
\end{eqnarray}
Notice that the list of other mixed relative correlation functions concerning
the nonextremal $D_1$-$D_5$-$P$-$\overline{P}$ black holes read as
\begin{eqnarray}
\frac{g_{ik}}{g_{il}}&=& \sqrt{\frac{n_l}{n_k}}, \ 
\frac{g_{ik}}{g_{jl}}= \frac{n_j}{n_i} \sqrt{\frac{n_l}{n_k}}, \ 
\frac{g_{kl}}{g_{ij}}= 0 \nonumber \\
\frac{g_{kl}}{g_{ii}}&=& 0, \ 
\frac{g_{kk}}{g_{ll}}= (\frac{n_l}{n_k})^{3/2}, \ 
\frac{g_{kl}}{g_{kk}}= 0
\end{eqnarray}

\section{Correlations for Six Charge Configurations}

Over generic nonlarge charge domains, we find from the entropy 
Eqn.(\ref{entropy6ne}) that the components of the covariant 
state-space metric tensor are given by the following expressions 
\begin{eqnarray}
g_{n_1 n_1}&=& \frac{\pi}{2n_1^{3/2}} ( \sqrt{n_2} + \sqrt{m_2} )( \sqrt{n_3}+ \sqrt{m_3} ), \ 
g_{n_1 m_1}= 0 \nonumber \\
g_{n_1 n_2}&=& -\frac{\pi}{2 \sqrt{n_1 n_2}} ( \sqrt{n_3}+ \sqrt{m_3} ), \ 
g_{n_1 m_2}= -\frac{\pi}{2 \sqrt{n_1 m_2}} ( \sqrt{n_3}+ \sqrt{m_3} ) \nonumber \\
g_{n_1 n_3}&=& -\frac{\pi}{2 \sqrt{n_1 n_3}} ( \sqrt{n_2}+ \sqrt{m_2} ), \ 
g_{n_1 m_3}= -\frac{\pi}{2 \sqrt{n_1 m_3}} ( \sqrt{n_2}+ \sqrt{m_2} ) \nonumber \\
g_{m_1 m_1}&=& \frac{\pi}{2m_1^{3/2}} ( \sqrt{n_2} + \sqrt{m_2} )( \sqrt{n_3}+ \sqrt{m_3} ), \ 
g_{m_1 n_2}= -\frac{\pi}{2 \sqrt{m_1 n_2}} ( \sqrt{n_3}+ \sqrt{m_3} ) \nonumber \\
g_{m_1 m_2}&=& -\frac{\pi}{2 \sqrt{m_1 m_2}} ( \sqrt{n_3}+ \sqrt{m_3} ), \ 
g_{m_1 n_3}= -\frac{\pi}{2 \sqrt{m_1 n_3}} ( \sqrt{n_2}+ \sqrt{m_2} ) \nonumber \\
g_{m_1 m_3}&=& -\frac{\pi}{2 \sqrt{m_1 m_3}} ( \sqrt{n_2}+ \sqrt{m_2} ), \ 
g_{n_2 n_2}= \frac{\pi}{2n_2^{3/2}} ( \sqrt{n_1} + \sqrt{m_1} )( \sqrt{n_3}+ \sqrt{m_3} ) \nonumber \\
g_{n_2 m_2}&=& 0, \ 
g_{n_2 n_3}= -\frac{\pi}{2 \sqrt{n_2 n_3}} ( \sqrt{n_1}+ \sqrt{m_1} ) \nonumber \\
g_{n_2 m_3}&=& -\frac{\pi}{2 \sqrt{n_2 m_3}} ( \sqrt{n_1}+ \sqrt{m_1} ), \ 
g_{m_2 m_2}= \frac{\pi}{2m_2^{3/2}} ( \sqrt{n_1} + \sqrt{m_1} )( \sqrt{n_3}+ \sqrt{m_3} ) \nonumber \\
g_{m_2 n_3}&=& -\frac{\pi}{2 \sqrt{m_2 n_3}} ( \sqrt{n_1}+ \sqrt{m_1} ), \ 
g_{m_2 m_3}= -\frac{\pi}{2 \sqrt{m_2 m_3}} ( \sqrt{n_1}+ \sqrt{m_1} ) \nonumber \\
g_{n_3 n_3}&=& \frac{\pi}{2n_3^{3/2}} ( \sqrt{n_1} + \sqrt{m_1} )( \sqrt{n_2}+ \sqrt{m_2} ), \ 
g_{n_3 m_3}= 0 \nonumber \\
g_{m_3 m_3}&=& \frac{\pi}{2m_3^{3/2}} ( \sqrt{n_1} + \sqrt{m_1} )(
\sqrt{n_2}+ \sqrt{m_2} )
\end{eqnarray}

In this case, from the definition of the relative statistical correlation functions,
for $i, j \in \{n_1, m_1 \}$, and $k, l \in \{n_2, m_2\}$, the relative correlation
functions satisfy the following scaling relations
\begin{eqnarray}
\frac{g_{ii}}{g_{jj}}&=& (\frac{j}{i})^{3/2}, \ 
\frac{g_{ii}}{g_{kk}}= (\frac{k}{i})^{3/2} (\frac{\sqrt{n_2}+\sqrt{m_2}}{\sqrt{n_3}+\sqrt{m_3}}), \ 
\frac{g_{ij}}{g_{ii}}= 0 \nonumber \\
\frac{g_{ii}}{g_{ik}}&=& -\frac{\sqrt{k}}{i} (\sqrt{n_2}+\sqrt{m_2}), \ 
\frac{g_{ik}}{g_{jk}}=  \sqrt{\frac{j}{i}}, \ 
\frac{g_{ii}}{g_{jk}}=  -\frac{\sqrt{jk}}{i^{3/2}} (\sqrt{n_2}+\sqrt{m_2}) \nonumber \\
\frac{g_{kk}}{g_{ik}}&=& -\frac{\sqrt{i}}{k} (\sqrt{n_2}+\sqrt{m_2}), \ 
\frac{g_{ij}}{g_{ik}}= 0, \ 
\frac{g_{ij}}{g_{kk}}= 0
\end{eqnarray}
The other concerned relative correlation functions are
\begin{eqnarray}
\frac{g_{ik}}{g_{il}}&=& \sqrt{\frac{l}{k}}, \ 
\frac{g_{ik}}{g_{jl}}= \sqrt{\frac{jl}{ik}}, \ 
\frac{g_{ij}}{g_{kl}}= n.d. \nonumber \\
\frac{g_{kl}}{g_{ii}}&=& 0, \ 
\frac{g_{kk}}{g_{ll}}= (\frac{l}{k})^{3/2}, \ 
\frac{g_{kl}}{g_{kk}}= 0
\end{eqnarray}
For $k, l \in \{n_2,m_2\}$, and $m, n \in \{n_3,m_3\}$, we have
\begin{eqnarray}
\frac{g_{kk}}{g_{mm}}&=& (\frac{m}{k})^{3/2} (\frac{\sqrt{n_3}+\sqrt{m_3}}{\sqrt{n_2}+\sqrt{m_2}}), \ 
\frac{g_{kl}}{g_{kk}}= 0, \ 
\frac{g_{kk}}{g_{km}}= -\frac{\sqrt{m}}{k} (\sqrt{n_3}+\sqrt{m_3})   \nonumber \\
\frac{g_{km}}{g_{lm}}&=& \sqrt{\frac{l}{k}}, \ 
\frac{g_{kk}}{g_{lm}}= -\frac{\sqrt{lm}}{k^{3/2}}(\sqrt{n_3}+\sqrt{m_3}), \ 
\frac{g_{mm}}{g_{km}}= -\frac{\sqrt{k}}{m} (\sqrt{n_2}+\sqrt{m_2}) \nonumber \\
\frac{g_{kl}}{g_{km}}&=& 0, \ 
\frac{g_{kl}}{g_{mm}}= 0
\end{eqnarray}
The other concerned relative correlation functions are
\begin{eqnarray}
\frac{g_{km}}{g_{kn}}&=& \sqrt{\frac{n}{m}}, \ 
\frac{g_{km}}{g_{ln}}= \sqrt{\frac{ln}{km}}, \ 
\frac{g_{kl}}{g_{mn}}= n. d. \nonumber \\
\frac{g_{mn}}{g_{kk}}&=& 0, \ 
\frac{g_{mm}}{g_{nn}}= (\frac{n}{m})^{3/2} , \ 
\frac{g_{mn}}{g_{mm}}= 0
\end{eqnarray}
Whilst, for $i, j \in \{n_1, m_1 \}$, and $m, n \in \{n_3,m_3\}$,
we have
\begin{eqnarray}
\frac{g_{ii}}{g_{mm}}&=& (\frac{m}{i})^{3/2} (\frac{\sqrt{n_3}+\sqrt{m_3}}{\sqrt{n_1}+\sqrt{m_1}}), \ 
\frac{g_{ij}}{g_{ii}}= 0, \ 
\frac{g_{ii}}{g_{im}}= -\frac{\sqrt{m}}{i} (\sqrt{n_3}+\sqrt{m_3})  \nonumber \\
\frac{g_{im}}{g_{jm}}&=& \sqrt{\frac{j}{i}} , \ 
\frac{g_{ii}}{g_{jm}}= -\frac{\sqrt{jm}}{i^{3/2}}(\sqrt{n_3}+\sqrt{m_3}), \ 
\frac{g_{mm}}{g_{im}}= -\frac{\sqrt{i}}{m} (\sqrt{n_1}+\sqrt{m_1}) \nonumber \\
\frac{g_{ij}}{g_{im}}&=& 0, \ 
\frac{g_{ij}}{g_{mm}}= 0, \ 
\frac{g_{im}}{g_{in}}= \sqrt{\frac{n}{m}} \nonumber \\
\frac{g_{im}}{g_{jn}}&=& \sqrt{\frac{jn}{im}}, \ 
\frac{g_{ij}}{g_{mn}}= n. d., \ 
\frac{g_{mn}}{g_{ii}}= 0, \ 
\frac{g_{mn}}{g_{mm}}= 0
\end{eqnarray}

\section{Principle Minors for Eight Charge Configurations}

For the entropy Eqn.(\ref{entropy8ne}) of the most general nonextremal 
nonlarge charged anticharged black hole in string involving finitely 
many nontrivially circularly fibered Kaluza-Klein monopoles,
the principle minors take the following expressions 
\ba p_1 &=& \frac{\pi}{2 n_1^{3/2}} (\sqrt{n_2}+\sqrt{m_2})
(\sqrt{n_3}+\sqrt{m_3}) (\sqrt{n_4}+\sqrt{m_4}), \nn
p_2&=& \frac{\pi^2}{4(n_1 m_1)^{3/2}} (\sqrt{n_2}+\sqrt{m_2})^2
(\sqrt{n_3}+\sqrt{m_3})^2 (\sqrt{n_4}+\sqrt{m_4})^2, \nn
p_3&=& \frac{\pi^3}{8(n_1 m_1 n_2)^{3/2}}
(\sqrt{n_3}+\sqrt{m_3})^3 (\sqrt{n_4}+\sqrt{m_4})^3
(\sqrt{n_2}+\sqrt{m_2}) \nn &&  \sqrt{m_2} (\sqrt{n_1}+\sqrt{m_1}), \nn
p_4&=&0, \nn
p_5&=& -\frac{\pi^5}{8(n_1 n_2 m_2 m_1)^{3/2} n_3}
(\sqrt{n_2}+\sqrt{m_2})^3 \nn &&  (\sqrt{n_3}+\sqrt{m_3})^3
(\sqrt{n_4}+\sqrt{m_4})^5 (\sqrt{n_1}+\sqrt{m_1})^3,\nn
p_6&=& -\frac{\pi^6}{16 (n_1 n_2 m_1 m_2 n_3 m_3)^{3/2}}
(\sqrt{n_2}+\sqrt{m_2})^4 \nn && (\sqrt{n_3}+\sqrt{m_3})^4
(\sqrt{n_4}+\sqrt{m_4})^6 (\sqrt{n_1}+\sqrt{m_1})^4,\nn
p_7&=& -\frac{\pi^7}{32 (n_1 m_1 n_2 m_2 n_3 m_3 n_4)^{3/2}}
(\sqrt{n_2}+\sqrt{m_2})^5 \nn && (\sqrt{n_3}+\sqrt{m_3})^5
(\sqrt{n_4}+\sqrt{m_4})^5 (4\sqrt{n_4}+\sqrt{m_4})\nn &&
(\sqrt{n_1}+\sqrt{m_1})^5,\nn
p_8&=& -\frac{\pi^8}{16 (\prod_{i=1}^4 n_i m_i)^{3/2}}
(\sqrt{n_2}+\sqrt{m_2})^6 \nn && (\sqrt{n_3}+\sqrt{m_3})^6
(\sqrt{n_4}+\sqrt{m_4})^6 (\sqrt{n_1}+\sqrt{m_1})^6.\ea


\begin{thebibliography}{99}

\bibitem{wein1} F. Weinhold, J. Chem. Phys. {\bf 63}, 2479 (1975).

\bibitem{wein2} F. Weinhold, J. Chem. Phys {\bf 63}, 2484 ( 1975).

\bibitem{rup1} G. Ruppeiner, Phys. Rev. {\bf A 20}, 1608
(1979).

\bibitem{rup11} G. Ruppeiner, Phys. Rev. Lett {\bf 50}, 287 (1983).

\bibitem{rup12} G. Ruppeiner, Phys. Rev. {\bf A 27}, 1116 (1983).

\bibitem{rup2} G. Ruppeiner, Rev. Mod. Phys {\bf 67} 605 (1995), Erratum {\bf
68},313 (1996).

\bibitem{rup21} G. Ruppeiner, C. Davis, Phys. Rev. {\bf A 41},
2200 (1990).

\bibitem{rup22}  G. Ruppeiner, Phys. Rev. {\bf D 75}, 024037 (2007).

\bibitem{RuppeinerPRD78} G. Ruppeiner,
Phy. Rev. D {\bf 78}, 024016 (2008).

\bibitem{bnt} B. N. Tiwari, {\tt arXiv:0801.4087v2 [hep-th]},  
\'Editions Universitaires Europ\'eennes, Germany (2011), ISBN: 978-613-1-53539-0,
New Paths Towards Quantum Gravity, Holbaek, Denmark (2008).

\bibitem{bnt1} T. Sarkar, G. Sengupta, B. N. Tiwari, J. High Energy Phys. {\bf 0810}, 076
(2008), {\tt arXiv:0806.3513v1 [hep-th]}.

\bibitem{bnt2} S. Bellucci, B. N. Tiwari, Entropy {\bf 12}, 2097  2010, {\tt arXiv:0808.3921 [hep-th]}.

\bibitem{bnt3} T. Sarkar, G. Sengupta, B. N. Tiwari, J. High Energy Phys. {\bf 0611}, 015 (2006), {\tt arXiv:hep-th/0606084}.

\bibitem{bnt5} S. Bellucci, B. N. Tiwari, Phys. Rev. D {\bf 82}, 084008,  (2010).

\bibitem{bnt6} S. Bellucci, B. N. Tiwari, J. High Energy Phys. {\bf 1011}, 030,  (2010), {\tt
arXiv:1009.0633v1 [hep-th]}.

\bibitem{bnt7} S. Bellucci, B. N. Tiwari, J. High Energy Phys. {\bf 1005}, 023,  (2010), {\tt arXiv:0910.5314v2
[hep-th]}.

\bibitem{bnt8} S. Bellucci, B. N. Tiwari, J. High Energy Phys. {\bf 1101}, 118, (2011), {\tt arXiv:1010.1427v1
[hep-th]}.

\bibitem{bnt9} S. Bellucci, B. N. Tiwari, Int. J. Mod. Phys. A  {\bf 26}, 32 (2011) 5403, {\tt arXiv:1010.3832v1
[hep-th]}.

\bibitem{bnt10} S. Bellucci, B. N. Tiwari, [Communicated], {\tt arXiv:1102.2391v1 [hep-th]}.

\bibitem{aman} J. E. Aman, I. Bengtsson, N. Pidokrajt, Gen. Rel. Grav. {\bf 38}, 1305, (2006)
{\tt gr-qc/0601119}.

\bibitem{aman1}  J. E. Aman, I. Bengtsson, N. Pidokrajt, Gen. Rel. Grav. {\bf 35}, 1733, (2003) {\tt
gr-qc/0304015}.

\bibitem{aman2} J. E. Aman, N. Pidokrajt, Phys. Rev. D {\bf 73}, 024017, (2006)
{\tt hep-th/0510139}.

\bibitem{aman2a} J. E. Aman, J. Bedford, D. Grumiller, N. Pidokrajt, J. Ward,
J. Phys. Conf. Ser. {\bf 66}, 012007, (2007), {\tt arxiv: gr-qc/0611119}.

\bibitem{aman3} G. Arcioni, E. Lozano-Tellechea, Phys. Rev. D {\bf 72}, 104021,
(2005) {\tt hep-th/ 0412118}.

\bibitem{aman4} J. Y. Shen, R. G. Cai, B. Wang, R.
K. Su,  Int. J. Mod. Phys. A {\bf 22}, 11, (2007),  {\tt gr-qc/0512035}.

\bibitem{aman5} M. Santoro, A. S. Benight, {\tt math-ph/0507026 }.

\bibitem{bntoo} B. N. Tiwari, {\tt arXiv:0801.3402v1 [hep-th]}. New Paths Towards
Quantum Gravity, Holbaek, Denmark (2008).

\bibitem{bntsbvc} S. Bellucci, V. Chandra, B. N. Tiwari, Int. J.
Mod. Phys. A {\bf 26}, 43, (2011), {\tt arXiv:0812.3792
[hep-th]}.

\bibitem{bullquark} S. Bellucci, V. Chandra, B. N. Tiwari,
Int. J. Mod. Phys. A {\bf 26}, 2665, (2011), {\tt arXiv:1010.4225v1 [hep-th]}.

\bibitem{Witten} E. Witten, Nucl. Phys. B {\bf 443}, 85, (1995).

\bibitem{9508072v3} S. Ferrara, R. Kallosh, A. Strominger,
Phys. Rev. D {\bf 52}, R5412,  (1995) {\tt
arXiv:hep-th/9508072v3}.

\bibitem{9602111v3} A. Strominger,
Phys. Lett. B {\bf 383} 39,  (1996), {\tt
arXiv:hep-th/9602111v3}.

\bibitem{new1} S. Ferrara, R. Kallosh,
Phys. Rev. D {\bf 54}, 1514,  (1996), {\tt
arXiv:hep-th/9602136}.

\bibitem{new2} S. Ferrara, G. W. Gibbons, R. Kallosh,
Nucl. Phys. B {\bf 500}, 75,  (1997), {\tt arXiv:hep-th/9702103}.

\bibitem{bfm1} S. Bellucci, S. Ferrara and A. Marrani, On some properties of the
Attractor Equations, Phys. Lett. B {\bf 635}, 172,  (2006), {\tt
hep-th/0602161}.

\bibitem{bfm2} S. Bellucci, S. Ferrara and A. Marrani, Supersymmetric Mechanics.
Vol.2: The Attractor Mechanism and Space-Time Singularities,
Springer-Verlag, Heidelberg, Lect. Notes Phys. {\bf 701}, 1-225, (2006).

\bibitem{bfgm1} S. Bellucci, S. Ferrara, M. G\"unaydin, A. Marrani,
Int. J. Mod. Phys. A {\bf 21}, 5043,  (2006), {\tt
hep-th/0606209}.

\bibitem{bfmy} S. Bellucci, S. Ferrara, A.
Marrani, A. Yeranyan, Riv. Nuovo Cim. {\bf 29N5} 1-88,  (2006), {\tt
hep-th/0608091}.

\bibitem{0702019v1} S. Bellucci, S. Ferrara, A.
Marrani, Contribution to the Proceedings of the XVII SIGRAV
Conference, 4-7 September 2006, Turin, Italy, {\tt
arXiv:hep-th/0702019}.

\bibitem{bskm} S. Bellucci, S. Ferrara, R. Kallosh, A. Marrani, Springer-Verlag, Heidelberg,
Lect. Notes Phys. {\bf 755}, 115,  (2008), {\tt arXiv:0711.4547v1
[hep-th]}.

\bibitem{08051310}  S. Bellucci, S. Ferrara, A. Marrani,
Fortsch. Phys. {\bf 56}, 761,  (2008), {\tt arXiv:0805.1310}.

\bibitem{bfgm2} S. Bellucci, S. Ferrara, M. G\"unaydin, A. Marrani,
SAM Lectures on Extremal Black Holes in d=4 Extended Supergravity,
Springer Proceedings in Phys. {\bf 134}, 1,  (2010), {\tt
arXiv:0905.3739 [hep-th]}.

\bibitem{0505122v2} A. Sen,
JHEP {\bf 0507}, 073, (2005), {\tt arXiv:hep-th/0505122v2}.

\bibitem{0411255} A. Sen,
JHEP {\bf 0505}, 059, (2005), {\tt arXiv:hep-th/0411255}.

\bibitem{0409148} A. Dabholkar,
Phys. Rev. Lett. {\bf 2005}, 94, 241301, (2005) {\tt
arXiv:hep-th/0409148}.

\bibitem{0507014v1} A. Dabholkar, F. Denef, G. W. Moore, B. Pioline,
JHEP {\bf 0510}, 096,  (2005), {\tt arXiv:hep-th/0507014v1}.

\bibitem{0502157v4} A. Dabholkar, F. Denef, G. W. Moore, B. Pioline,
JHEP {\bf 0508}, 021, (2005), {\tt arXiv:hep-th/0502157v4}.

\bibitem{ref1} J. de Boer, M. Shigemori ,
``Exotic branes and non-geometric backgrounds",
Phys. Rev. Lett. {\bf 104}, 251603, (2010),
{\tt arXiv:1004.2521 [hep-th]}.


\bibitem{0504005} A. Sen,
Adv. Theor. Math. Phys. {\bf 9}, 527, 2005, {\tt
arXiv:hep-th/0504005}.

\bibitem{0611330} B. D. Chowdhury, S. D. Mathur, {\tt
arXiv:hep-th/0611330}.


\bibitem{GarousiGhodsiCai} M. R. Garousi, A. Ghodsi,
JHEP {\bf 0710},  036, (2007), {\tt arXiv:0705.2149v2 [hep-th]}.

\bibitem{GarousiGhodsiCai1} R. G. Cai, D. W. Pang,
JHEP {\bf 0705}, 023, (2007), {\tt arXiv:hep-th/0701158v2}.

\bibitem{GarousiGhodsiCai2} M. R. Garousi, A. Ghodsi,
JHEP {\bf 043} (2007) 04, {\tt arXiv:hep-th/0703260}.



\bibitem{maldacena} J. M. Maldacena, Adv. Theor. Math. Phys. {\bf 2},
231, (1998).

\bibitem{0412322} D. Gaiotto, A. Strominger and X. Yin:
JHEP {\bf 0511}, 017, (2005), {\tt arXiv:hep-th/0412322}.

\bibitem{hidden1} A. Castro, A. Maloney, A. Strominger,
{\tt arXiv:1004.0996v1 [hep-th]}

\bibitem{hidden2} T. Hartman, K. Murata, T. Nishioka, A. Strominger, JHEP {\bf 0904}
(2009) 019.


\bibitem{grqc9307038} R. M. Wald,
Phys. Rev. D, {\bf 48}, R3427, (1993), {\tt
arXiv:gr-qc/9307038}.

\bibitem{grqc9502009} T. Jacobson, G. Kang, R. C. Myers,
McGill, {\bf 95} 04; UMDGR, {\bf 95}, 092, {\tt
arXiv:gr-qc/9502009}.

\bibitem{grqc9912119} R. M. Wald, Living Rev. Rel. {\bf 4}, 6, (2001), {\tt
arXiv: gr-qc/9912119}.

\bibitem{9305016} T. Jacobson, R. C. Myers,
Phys. Rev. Lett. {\bf 70}, 3684, (1993), {\tt
arXiv:hep-th/9305016}.


\bibitem{9512078v1} C. Vafa, Nucl. Phys. B, {\bf 463},
435, (1996), {\tt arXiv:hep-th/9512078v1}.

\bibitem{9601029v2} A. Strominger, C. Vafa,
Phys. Lett. B, {\bf 379}, 99, (1996), {\tt
arXiv:hep-th/9601029v2}.

\bibitem{MSW} J. M. Maldacena, A. Strominger, E. Witten,
JHEP {\bf 9712}, 002, (1997),
{\tt arXiv:hep-th/9711053}.

\bibitem{9602043v2} C. G. Callan, J. M. Maldacena,
Nucl. Phys. B, {\bf 472}, 591, (1996), {\tt
arXiv:hep-th/9602043v2}.

\bibitem{9603109v1} G. Horowitz, J. M. Maldacena, A. Strominger,
Phys. Lett. B, {\bf 383}, 151, (1996), {\tt
arXiv:hep-th/9603109v1}.

\bibitem{9603195v1} G. T. Horowitz, D. A. Lowe, J. M. Maldacena,
Phys. Rev. Lett. {\bf 77}, 430, (1996), {\tt
arXiv:hep-th/9603195v1}.

\bibitem{9603061v2} C. V. Johnson, R. R. Khuri, R. C. Myers,
Phys. Lett. B, {\bf 378}, 78, (1996), {\tt
arXiv:hep-th/9603061v2}.


\bibitem{07052564v1} F. Denef, G. W. Moore,
{\tt arXiv:0705.2564v1 [hep-th]}.

\bibitem{0702146v2} F. Denef, G. W. Moore,
{\tt arXiv:hep-th/0702146v2}.




\bibitem{0604110} I. Bena, C.W. Wang, N. P. Warner,
NSF-KITP-06-25, CERN-PH-TH/2006-045,
{\tt arXiv:hep-th/0604110v2}.

\bibitem{07063786V2} I. Bena, C. W. Wang, N. P. Warner,
JHEP  {\bf 0807}, 019, (2008), {\tt arXiv:0706.3786v2 [hep-th]}.



\bibitem{hepth0508023} V. Balasubramanian, J. de Boer, V. Jejjala, J. Simon,
 JHEP {\bf 0512}, 006, (2005), {\tt arXiv:hep-th/0508023v2}.

\bibitem{hepth0107119} V. Balasubramanian, M. Berkooz, A. Naqvi, and M. J. Strassler,
JHEP {\bf 0204}, 034, (2002), {\tt arXiv:hep-th/0107119}.


\bibitem{GRG} R. Emparan, H. S. Reall,
Gen. Rel. Grav. {\bf 34}, 2057, (2002). 

\bibitem{HoMy} J. L. Hovdebo, R. C. Myers,
  Phys. Rev. D {\bf 73}, 084013, (2006),
 {\tt arXiv:hep-th/0601079}.

\bibitem{EEV} H. Elvang, R. Emparan, A. Virmani,
{\tt arXiv:hep-th/0608076}.

\bibitem{GL} R. Gregory, R. Laflamme,
  Phys. Rev. Lett. {\bf 70}, 2837, (1993),
 {\tt arXiv:hep-th/9301052}.

\bibitem{LoTeAr} G. Arcioni, E. L. Tellechea,
  Phys. Rev. D {\bf 72}, 104021, (2005), {\tt arXiv:hep-th/0412118}.

\bibitem{odias} O. J. C. Dias,
  Phys. Rev. D {\bf 73}, 124035, (2006), {\tt arXiv:hep-th/0602064}.


\bibitem{LLM} H. Lin, O. Lunin, and J. Maldacena,
JHEP {\bf 0410}, 025 (2004).


\bibitem{0109154v1} O. Lunin, S. D. Mathur,
Nucl. Phys. B  {\bf 623}, 342-394, (2002), {\tt arXiv:hep-th/0109154v1}.

\bibitem{0202072v2} O. Lunin, S. D. Mathur,
Phys. Rev. Lett.  {\bf 88}, 211303, (2002), {\tt
arXiv:hep-th/0202072v2}.

\bibitem{0706.3884v1} S. D. Mathur,
{\tt arXiv:0706.3884v1 [hep-th] }.

\bibitem{0804.0552}  K. Skenderis and M. Taylor,
  Phys. Rept.  {\bf 467}, 117-171, (2008),
 {\tt  arXiv:0804.0552 [hep-th]}.


\end{thebibliography}
\end{document}